\documentclass[11pt]{article}

\usepackage{epsfig}
\usepackage{color}

\usepackage{graphicx}

\newcommand{\be}{\begin{equation}}
\newcommand{\ee}{\end{equation}}
\newcommand{\bea}{\begin{array}}
\newcommand{\ea}{\end{array}}
\newcommand{\beqa}{\begin{eqnarray}}
\newcommand{\eeqa}{\end{eqnarray}}
\newcommand{\bean}{\begin{eqnarray*}}
\newcommand{\eean}{\end{eqnarray*}}

\newcommand{\gapproxeq}{\lower .7ex\hbox{$\;\stackrel{\textstyle
>}{\sim}\;$}}
\newcommand{\lapproxeq}{\lower .7ex\hbox{$\;\stackrel{\textstyle
<}{\sim}\;$}}

\newcounter{appendice}

\def\thebibliography#1{{\bf REFERENCES\markboth
 {REFERENCES}{REFERENCES}}\list
 {[\arabic{enumi}]}{\settowidth\labelwidth{[#1]}\leftmargin\labelwidth
 \advance\leftmargin\labelsep
 \usecounter{enumi}}
 \def\newblock{\hskip .11em plus .33em minus -.07em}
 \sloppy
 \sfcode`\.=1000\relax}

\def\up#1{\leavevmode \raise.16ex\hbox{#1}}

\def\sqr#1#2{{\vcenter{\vbox{\hrule height.#2pt
        \hbox{\vrule width.#2pt height#1pt \kern#1pt
          \vrule width.#2pt}
        \hrule height.#2pt}}}}

\begin{document}

\centerline{ \LARGE Location and Direction Dependent Effects }
\vskip .5cm
\centerline{ \LARGE in Collider Physics from  Noncommutativity}

\vskip 2cm

\centerline{   Mansour Haghighat$^{a,b}$\footnote{mansour@cc.iut.ac.ir}, Nobuchika Okada$^{a}$\footnote{okadan@ua.edu} and Allen Stern$^{a}$\footnote{astern@bama.ua.edu}   }

\vskip 1cm
\begin{center}
  {\it a) Department of Physics, University of Alabama,\\ Tuscaloosa,
Alabama 35487, USA\\}
{\it b) Department of Physics, Isfahan University of Technology,\\ Isfahan 84156-83111, Iran\\}
\end{center}
\vskip 2cm

\vspace*{5mm}

\normalsize
\centerline{\bf ABSTRACT}
We examine the leading order noncommutative corrections to the differential and total cross sections for $e^+e^- \rightarrow q\;\bar q$. After averaging over the earth's rotation, the results depend on the latitude for the collider,  as well as the direction of the incoming beam. They also depend on scale and direction of the noncommutativity. Using data from LEP, we  exclude regions in the parameter space spanned by the noncommutative scale and angle relative to the earth's axis.  We also investigate possible implications for phenomenology at the future International Linear Collider.  
  
\newpage

\section{Introduction}
\setcounter{equation}{0}
Motivated in part by quantum gravity \cite{Doplicher:1994zv}, it has
been of recent interest to examine field theories, and in particular  the
standard model of particle physics, on  noncommutative space-time
backgrounds. Noncommutative versions of the standard model have been
proposed, which have the potential of explaining its gauge algebra and
fermion representations \cite{Chamseddine:2007hz}.
Two popular noncommutative  generalizations of the standard model were given by Chaichian et. al. \cite{Chaichian:2001py} and Calmet et. al. \cite{Calmet:2001na}. In both cases the geometry is generated by  Heisenberg algebras, i.e.,
\be [{\bf x}^\mu,{\bf x}^\nu ]=i\Theta^{\mu\nu} \;,\qquad\ee where ${\bf x}^\mu,\;\mu,\nu,...=0,1,2,3$, are operator analogues of space-time coordinates  and 
$\Theta^{\mu\nu}=-\Theta^{\nu\mu}$ are central elements which are independent of ${\bf x}^\mu$.  Also, both approaches rely on the Moyal-Weyl star product realization of the algebra.
One often distinguishes two cases: space-space noncommutativity  associated with $\Theta^{ij}$, $ i,j,k,...=1,2,3,$ and
time-space noncommutativity  associated with $\Theta^{0i}$. Then two noncommutative energy scales $\Lambda_{SS}$, $\Lambda_{TS}$, and two unit
vectors $v_i$,  $w_i$ can be defined using
\be 
\Theta^{ij}=\frac{1}{2\Lambda_{SS}^2}\epsilon_{ijk}{v_{k}}\;\qquad \Theta^{0i}=\frac 1{\Lambda_{TS}^2} {w_{i}}\;.\label{(1.3)}\ee
Bounds on $\Lambda_{SS}$ and $\Lambda_{TS}$
 have been obtained from  atomic physics,  collider physics and astrophysics.  (See for example \cite{bonds}.)
$v_i$ and  $w_i$ correspond to fixed directions in space.  If
the noncommutative scales become accessible in collider physics, then information about these directions may be obtainable, as we shall illustrate in this article. 

We first make some brief remarks about the two approaches  to the noncommutative standard model mentioned above. In the approach of  \cite{Chaichian:2001py}, one enlarges the standard model gauge  group to the noncommutative analogue of $U(3)\times U(2)\times U(1)$, thus introducing gauge bosons in addition to those of the standard model.  New symmetry breakings and Higgs scalars are then also required.  The model was shown to be one-loop renormalizable\cite{bonora}.  The approach of
 \cite{Calmet:2001na}  does not involve introducing any additional gauge bosons or
symmetry breakings.   It instead relies upon a map, known as the Seiberg-Witten map\cite{Seiberg:1999vs}, between commutative and noncommutative gauge theories, from which one can obtain corrections to the
standard model interactions \cite{Melic:2005fm}.  These corrections
have been computed up to second order in  $\Theta^{\mu\nu}$ \cite{Alboteanu:2007bp}.
In contrast with the former model,   one-loop corrections  are not well understood in this approach.  Although  the model is
anomaly free up to one-loop order \cite{martin}, only the pure gauge sector of this theory has been shown to be renormalizable  at this order \cite{renorm-exnc}.  
 
 Here we shall follow the  approach of \cite{Calmet:2001na}, and obtain all  leading noncommutative corrections to the differential and total cross sections for the example of $e^+e^- \rightarrow q\;\bar q$ at tree level. These corrections are  second order in  $\Theta^{0i}$. Assuming that $w_{i}$ in (\ref{(1.3)}) corresponds to a fixed direction relative to some frame external to the earth, and not the lab frame, we must  average over the earth's rotation. (Presumably, other effects due to the earth's motion relative to $w_{i}$ are much smaller.)  Results for the cross sections then depend on the latitude for the collider,  as well as the direction of the incoming beam.  We can obtain an analytic expression for the averaged leading order noncommutative correction to the total cross section in terms of these quantities, along with the noncommutative scale $\Lambda_{TS}$ and the projection $w_Z$ of $w_i$ along the earth's axis. 
 Using data from LEP, we are then able to exclude regions in the parameter space spanned by $\Lambda_{TS}$ and  $w_Z$ for different detectors.  Here we also investigate possible implications of the noncommutative corrections for phenomenology at the International Linear Collider (ILC), 
  a future high energy $e^+ e^-$ linear collider 
with $\sqrt{s}=$500 GeV$-$1 TeV. 
Since the corrections depend on both the location 
of the  ILC and its beam direction, there can be an optimal site and beam direction  for observing  noncommutative effects. 
 
 Before proceeding, we comment on several works where noncommutative corrections for the annihilation of $e^+e^-$ to fermion-antifermion pairs, and where earth rotational effects in collider physics, were already considered.  $e^+e^- \rightarrow e^+e^-$ was examined  in \cite{Hewett:2000zp} in a version of noncommutative QED that did not rely on the Seiberg-Witten map.   The calculations included Z-boson exchange, although the Z-boson vertex was not obtained  from a noncommutative standard model lagrangian.  Electron-positron scattering
  was reconsidered in   \cite{Das:2007dn} within the framework of the noncommutative standard model \cite{Calmet:2001na}, which gives a specific  form for
 the Z-boson vertex.  Electron scattering was examined in another approach to noncommutative QED in \cite{Balachandran:2006ib}.    Earth rotation effects were not taken into account in these works.
 Such  effects were considered for collider physics in \cite{Grosse:2001xz}, using the example of Higgs pair production.  Earth rotation effects were also illustrated in \cite{Kamoshita:2002wq} for $e^-e^+\rightarrow \gamma\gamma$ and 
 in \cite{liao:2003} for  $e^+e^- \rightarrow \mu^+\mu^-$.  The investigations in  \cite{Grosse:2001xz},\cite{Kamoshita:2002wq},\cite{liao:2003} were conducted   within the context of noncommutative  QED, and  Z-boson exchange was not included.  Scattering amplitudes were only expanded up to first  order in $\Theta^{\mu\nu}$ in the above mentioned mentioned works, although  the leading order corrections to the cross section for $e^+e^-$ to fermion-antifermion pairs are quadratic.  So additional terms can contribute to the cross section at this order. In our work, upon applying  the noncommutative standard model of \cite{Calmet:2001na}, and not just noncommutative QED, we shall include all second order contributions to  the scattering amplitude,  in addition to taking into account earth rotational effects.

 The outline for the rest of this article is as follows:  In section 2 we compute the noncommutative corrections for  photon and Z-boson exchange.  We average over the earth's rotation in section 3 and apply the results to LEP and  ILC in section 4.

\section{Application of the Noncommutative Standard Model}
\setcounter{equation}{0}

We  first give the noncommutative Feynman rules for the relevant vertices $\gamma ff$ and  $Zff$ up to second order in $\Theta^{\mu\nu}$.  There are no noncommutative corrections to the propagators when one uses the Moyal-Weyl star product.

The Feynman rule for vertex $\gamma ff$ is given by 
%
\be 
V^{\gamma ff}_{\mu}=i  e  Q_{f}  \Gamma_{\mu}\;,\ee
where $Q_f$  denotes the fermion charge and one can expand $\Gamma_{\mu}$ in a power series in the noncommutativity tensor $\Theta^{\mu\nu}$,
\be  \Gamma_{\mu}=\gamma_\mu + \Gamma^{(1)}_{\mu}+ \Gamma^{(2)}_{\mu}+\cdots\;\ee Here, and in what follows, the dots denote terms that are more than second order in $\Theta^{\mu\nu}$.
The first and second order terms were found in \cite{Melic:2005fm} and \cite{Alboteanu:2007bp}, respectively, to be
\beqa
\Gamma^{(1)}_{\mu}&=&
 \frac{i}{2}
\left[
(k\Theta)_\mu\slash \!\!\! p_{\mbox{\tiny in}}(1- 4c^{(1)}_\psi)+ 2(k\Theta)_\mu\slash \!\!\!k(c^{(1)}_A-c^{(1)}_\psi)
-( p_{\mbox{\tiny in}}\Theta)_\mu
\slash \! \! \! k-
(k \Theta p_{\mbox{\tiny in}}) \gamma_\mu
\right]\cr &&\cr
\Gamma^{(2)}_{\mu}&=& \frac{1}{8}(k\Theta p_{\mbox{\tiny in}})
\left[
(k\Theta)_\mu\slash \!\!\! p_{\mbox{\tiny in}}(1- 16c^{(2)}_\psi)+ 4(k\Theta)_\mu\slash \!\!\!k(c^{(1)}_A-2c^{(2)}_\psi)
-( p_{\mbox{\tiny in}}\Theta)_\mu
\slash \! \! \! k-
(k \Theta p_{\mbox{\tiny in}}) \gamma_\mu
\right]\;,\cr &&
\label{eq:ffgamma}
\eeqa
where  we  ignore the fermion mass and taken all  momenta, $k$, $p_{\mbox{\tiny in}}$ and $p_{\mbox{\tiny out}}$, to be incoming.
$p_{in}$ and $p_{out}$ are associated with incoming and outgoing fermions, respectively. We have adopted the notation $(k\Theta)_\mu=k^\nu\Theta_{\nu\mu}$ and $(k\Theta p)=k^\nu\Theta_{\nu\mu}p^\mu $. $c^{(1)}_\psi$, $c^{(1)}_A$, and $c^{(2)}_\psi$ are arbitrary constants which originate from ambiguities in the Seiberg-Witten map.  They do not appear in the vertex when the incoming and outgoing fermions are evaluated on shell (at energies well above the fermion mass).  In that case, the vertex reduces to
\be
V^{\gamma ff}_{\mu} \Big|_{ {{\rm on-shell}}}=i  e  Q_{f}\gamma_\mu\;{\cal I}({p_{\mbox{\tiny out}}, p_{\mbox{\tiny in}})}\;,\label{gmaffoshl}\ee where 
\be {\cal I}({p_{\mbox{\tiny out}}, p_{\mbox{\tiny in}})}=1 +
 \frac{i}{2}
(p_{\mbox{\tiny out}} \Theta p_{\mbox{\tiny in}}) +
 \frac{1}{8}(p_{\mbox{\tiny out}}\Theta p_{\mbox{\tiny in}})^2
 +\cdots
\ee 
The sign in front of the second term  changes upon switching the ingoing momenta $k$ to outgoing.

The Feynman rule for vertex  $Zff$ 
%
%
%
is obtained by replacing 
$ Q_{f}  \gamma_{\mu}$ in (\ref{gmaffoshl}) by 
$ {(c_{V,f}-c_{A,f}\gamma_5)}
\gamma_{\mu}/{\sin2\theta_W}\;,$ 
where 
\beqa c_{V,f} &=&\frac12( c_{L,f}+c_{R,f})\;=\;T_{3,f} - 2 Q_f \sin^2\theta_W\cr
c_{A,f}&=& \frac12( c_{L,f}-c_{R,f})\;=\;T_{3,f}\;,\eeqa
$\theta_W$ is the Weinberg angle and $T_{3,f}$ denotes the fermion weak isospin. 
The on-shell vertex is then
\be
V^{Z ff}_{\mu} \Big|_{ {{\rm on-shell}}}=\frac{i e}{\sin2\theta_W}\;{(c_{V,f}-c_{A,f}\gamma_5)}
\gamma_{\mu}\;{\cal I}({p_{\mbox{\tiny out}}, p_{\mbox{\tiny in}})}\label{Zffoshl}
\ee

Up to second order in $\Theta^{\mu\nu}$, both noncommutative on shell vertices (\ref{gmaffoshl}) and (\ref{Zffoshl}) are related to the commutative on shell vertices by the same factor ${\cal I}({p_{\mbox{\tiny out}}, p_{\mbox{\tiny in}})}$.  It follows that noncommutative  scattering
amplitudes for $e^-e^+\rightarrow q \;\bar q$ associated with  $\gamma$ and $Z$ exchanges are  related
to their commutative  counterparts by a common factor.  Then the total noncommutative  scattering amplitude at tree level ${\cal{M}^{NC}}$ is related to  total commutative  scattering amplitude ${\cal{M}}$ by 
\be
{\cal{M}^{NC}}={\cal I}({p_, p_{\mbox{\tiny in}})}^*\;{\cal I}({p^\prime_{\mbox{\tiny out}}, p^\prime_{\mbox{\tiny in}})}\;{\cal M }\;,
\ee
where the primed momenta are associated with the created fermions, 
and ${\cal M}$ is the corresponding standard model amplitude.   
The leading noncommutative corrections to the  squared-amplitude are second order in $\Theta^{\mu\nu}$,
\be
\mid{\cal{M}^{NC}}\mid^2=\left[ 1+\frac{1}{2} (p_{\mbox{\tiny out}} \Theta p_{\mbox{\tiny in}})^2\,+\frac{1}{2} 
(p^\prime_{\mbox{\tiny out}} \Theta  p^\prime_{\mbox{\tiny in}})^2 \,+\cdots\right]\mid{\cal M}\mid^2.
\ee
Only the space-time components of $\Theta_{\mu\nu}$ contribute  in
center of mass frame for beam on beam scattering, where 
\be p_{\mbox{\tiny in}}= (\frac{\sqrt s}{2},\vec p)\;, \qquad  
 p_{\mbox{\tiny out}}= (\frac{\sqrt s}{2},-\vec p)\;,\ee
 and similarly,  
\be p'_{\mbox{\tiny in}}= (\frac{\sqrt s}{2},\vec p')\; , \qquad  
p'_{\mbox{\tiny out}}= (\frac{\sqrt s}{2},-\vec p').\ee 
Then  using (\ref{(1.3)}),\be
 (p_{\mbox{\tiny out}} \Theta p_{\mbox{\tiny in}})=\frac{\sqrt
{ s}}{\Lambda_{TS}^2}\;\overrightarrow{p}\cdot \vec w \; , \qquad (p'_{\mbox{\tiny out}} \Theta p'_{\mbox{\tiny in}})=\frac {\sqrt{ s}}{\Lambda_{TS}^2}\;\overrightarrow{p}'\cdot \vec w\;,\ee   where $\vec w=\{w_i\}$.
In terms of unit vectors $ \hat{p}=\vec p/|p|$ and $ \hat{p'}=\vec p'/|p'|$, one can then write the noncommutative differential cross section $d\sigma^{\cal{ NC}}/d\Omega$ for $e^+e^-\rightarrow q\; \bar q$ according to
\be
\frac{d\sigma^{\cal{ NC}}}{d\Omega}=\left[ 1+\frac18 \biggl(\frac{s}{\Lambda^2_{TS}}\biggr)^2\Bigl\{ (\hat{p}\cdot\vec  w)^2\,+(\hat{p}^\prime\cdot \vec w)^2 \Bigr\}+\cdots\right]\frac{d\sigma}{d\Omega}\;,\label{nccrsctn}
\ee
where   $d\sigma/d\Omega$ is the standard model differential cross section.
This expression is valid at lowest order in perturbation theory 
provided that $\Lambda_{TS}\; {{}^>_{{}^\sim} }\;{\sqrt s}$. The standard model
differential cross section  and total cross section $\sigma_{\rm total}$
are well known \cite{Burgess:2007zi}
\beqa
\frac{d\sigma}{d\Omega}&=&\frac{N_c\alpha^2 s}{16 }\biggl\{ F(s)(1+\cos \beta)^2 + G(s)(1-\cos \beta)^2 \biggr\} \; ,\label{smdfcrsn}
\\ &&\cr
\sigma_{\rm total}&=&\frac{N_c\alpha^2\pi s}{3 }\Bigl( F(s) + G(s)\Bigr)\;, \label{smttlcrsn}
\eeqa
where $\beta$ denotes the scattering angle, $N_c=3$ is the number of colors,
\be F(s) = |A_{LL}(s)|^2+ |A_{RR}(s)|^2\;,\qquad \quad G(s) = |A_{LR}(s)|^2+ |A_{RL}(s)|^2\;,\ee
\be
A_{ij}(s)= \frac{Q_eQ_f}s+\frac {c_{i,e}c_{j,f}}{\sin^2\theta_{W}\cos^2\theta_{W}}\;\frac 1{s-M_Z^2-iM_Z \Gamma_Z} \;,\qquad i,j = L,R\;,\ee and we have included the correction due to the decay width $\Gamma_Z$ for $Z$.

\section{Earth rotational effects}
\setcounter{equation}{0}

Now we take into account  earth rotational effects.   This is necessary since  $\hat{p}$ and $\hat{p}^\prime $ are defined in the lab frame, while $\vec w$ is a fixed direction in space, and so the earth's rotation implies that the scalar products appearing in (\ref{nccrsctn}) are not constant.  As was reported in \cite{Grosse:2001xz}, this  can lead to a day-night asymmetry in the cross section.  Since such time-dependent experimental  data are not readily available, we shall average $(\hat{p}\cdot\vec w)^2$ and $(\hat{p}^\prime\cdot\vec w)^2$ in (\ref{nccrsctn}) over a full day. 

Following \cite{Kostelecky:1999mr}, denote by $(\hat X,\hat Y,\hat Z)$ a nonrotating basis, with $\hat Z$ parallel to the earth's axis along the north direction.  To a good approximation, this basis spans an inertial frame.  The transformation to a basis $(\hat x,\hat y,\hat z)$ attached to a point on the earth's surface  at any time $t$ was given by
\be \pmatrix{ \hat x\cr \hat y\cr \hat z\cr}= \pmatrix{\cos \chi \cos {\Omega t}&
\cos \chi \sin {\Omega t}& -\sin \chi\cr-\sin {\Omega t} & \cos{\Omega t} &0\cr 
\sin \chi \cos {\Omega t}&
\sin \chi \sin {\Omega t}& \cos \chi\cr}\pmatrix{ \hat X\cr \hat Y\cr \hat Z\cr}\;,\ee
where $\Omega$ is the earth's sidereal frequency and $0\le\chi\le \pi$.  To identify the directions $(\hat x,\hat y,\hat z)$, consider the cases $\chi =0$ and $\chi =\pi/2$, which we identify with the north pole and equator, respectively.
$\hat z\parallel \hat Z $ when $\chi =0$, and therefore,  $\hat z$ points normal to the earth's surface.  $\hat x$ is anti-parallel to $ \hat Z $ when $\chi =\pi/2$, and thus $\hat x$ and $\hat y$ point south and east, respectively. 

Assuming no vertical component to the particle momentum $\vec p$ in the lab frame, we have \be \hat p = \cos\phi \;\hat x + \sin\phi\; \hat y  \ee
Taking $\vec w = w_X \hat X + w_Y \hat Y + w_Z \hat Z $, the time average of $(\vec w\cdot \hat p)^2$ is
\be <( \hat p\cdot\vec w )^2> \;=\; \frac 12 \left(\cos^2 \phi \cos^2 \chi +\sin^2 \phi\right)\left(1-w_Z ^2\right)\; +\; \cos^2 \phi \sin^2 \chi w_Z ^2\;,\ee
and so the average leading order correction to the standard model differential cross section  (\ref{smdfcrsn}) is
\beqa \Big< \delta \frac{d\sigma}{d\Omega}\Big> & =&
\biggl(\frac{s}{4\Lambda^2_{TS}}\biggr)^2\Bigl\{ \Bigl(\cos^2 \phi
\cos^2 \chi +\sin^2 \phi+\cos^2 (\phi+\beta) \cos^2 \chi +\sin^2
(\phi+\beta)\Bigr)\left(1-w_Z ^2\right)\;\cr & & \qquad+\;
2\Bigl(\cos^2 \phi +\cos^2 (\phi +\beta)\Bigr)\sin^2 \chi\; w_Z ^2
\Bigr\}\;\frac{d\sigma}{d\Omega} \label{crfrdfsgm}\eeqa
Note that this correction is always positive. 
Upon integrating this plus (\ref{smdfcrsn}) over the scattering angle, 
we obtain the following analytic formula for the  average leading order correction to the  total cross section (\ref{smttlcrsn}):
\beqa <\delta\sigma_{\rm total}> &=&\frac{  s^2 }{5120
   \Lambda_{TS} ^4}\; \Biggl\{15\pi r(s)
    \Bigl[\left(3 w_Z^2-1\right) \cos 2 \chi -(w_Z^2+1)\Bigr] \sin 2 \phi \cr & &\cr& &\quad+32 
   \Bigl[2 (w_Z^2+1) \cos 2 \phi -\Bigl(5( \cos 2 \chi+1) +3 \cos 2 (\phi +\chi )
   \Bigr)w_Z^2\cr & &\cr & &+(1-3 w_Z^2) \cos 2 (\phi -\chi )-5 \cos 2 \chi +\cos 2 (\phi +\chi )+15\Bigr]\Biggr\}\;\sigma_{\rm total}\;,\cr & &\label{ttlncsgm}\eeqa
   where \be r(s)=
   \frac{F(s)-G(s)}{F(s)+G(s)} \ee

\section{Numerical results for collider physics}
\setcounter{equation}{0}

We now apply the results from LEP.  
Our calculations  for the corrections to total cross section 
of $e^-e^+ \to q \; {\bar q}$ can be compared to measurements made at the four detectors 
ALEPH, DELPHI, OPAL and L3, which were spaced at $90^\circ$ 
degree intervals around the ring.  
Since (\ref{ttlncsgm}) is unchanged for 
$\phi\rightarrow \phi + \pi$,  only two distinct answers are obtained for the four of the detectors. 
Below we will apply the results of ALEPH, and  OPAL.

For $\sqrt{s}\simeq 189$ GeV, we get a contribution of $\sim 10.8$ pb  to the 
total standard model cross section $\sigma_{\rm total}$ from $q=u,c$, and 
$\sim 10.1$ pb from $q=d,s,b$, 
where we have cut the region $|\cos \beta| > 0.95$ from the integration 
corresponding to data analysis of LEP experiments. 
The latitude for CERN is $46.234^\circ$, corresponding to $\chi=0.764$ radians.  
Upon taking $\phi \simeq -\pi/ 3$ for the ALEPH detector, 
and summing over all five quark final states,  we find 
the deviation of total cross section from the standard model one as
\be 
\frac{\sum_{quarks}<\delta\sigma_{\rm total}>|_{{}_{\mbox{\tiny
ALEPH}}}}{\sum_{quarks}\sigma_{\rm total}}\; 
\approx\Big(\frac{104.6\;{\rm GeV}}{\Lambda_{TS}}\Bigr)^4 
- \Big(\frac{79.5\;{\rm GeV}}{\Lambda_{TS}}\Bigr)^4w_Z^2 \; .
\label{ALEPH}
\ee
Similarly, if we take  $\phi=-5\pi/ 6$ for the OPAL detector, we find
\be 
\frac{\sum_{quarks}<\delta\sigma_{\rm total}>|_{{}_{\mbox{\tiny
OPAL}}}}{\sum_{quarks}\sigma_{\rm total}}\; 
\approx\Big(\frac{105.3\;{\rm GeV}}{\Lambda_{TS}}\Bigr)^4 
- \Big(\frac{84.1\;{\rm GeV}}{\Lambda_{TS}}\Bigr)^4w_Z^2 \; .
\label{OPAL}
\ee
We set (\ref{ALEPH}) equal to the error found 
in the ALEPH results of $3.74$\% \cite{ALEPH} and (\ref{OPAL}) 
equal to the error found in the OPAL results of $1.35$\% \cite{OPAL}%
\footnote{
Here, we refer to the ALEPH results for $\sqrt{s'/s} >0.85$  
and the OPAL results for $\sqrt{s'/s} >0.7225$, 
where $s'$ is the effective center-of-mass energy 
of $e^+e^-$ collisions. 
Although the initial state radiations are not taken into account 
in our analysis, they are not significant for such high cuts 
on $\sqrt{s'/s}$. 
}. 
%
%
Fig.~1 shows contour plots for ALEPH (dashed) and OPAL (solid)
results, respectively. 
The left hand side of each of the contours is excluded.

\begin{figure}[t]\begin{center}
\includegraphics[scale=1.2]{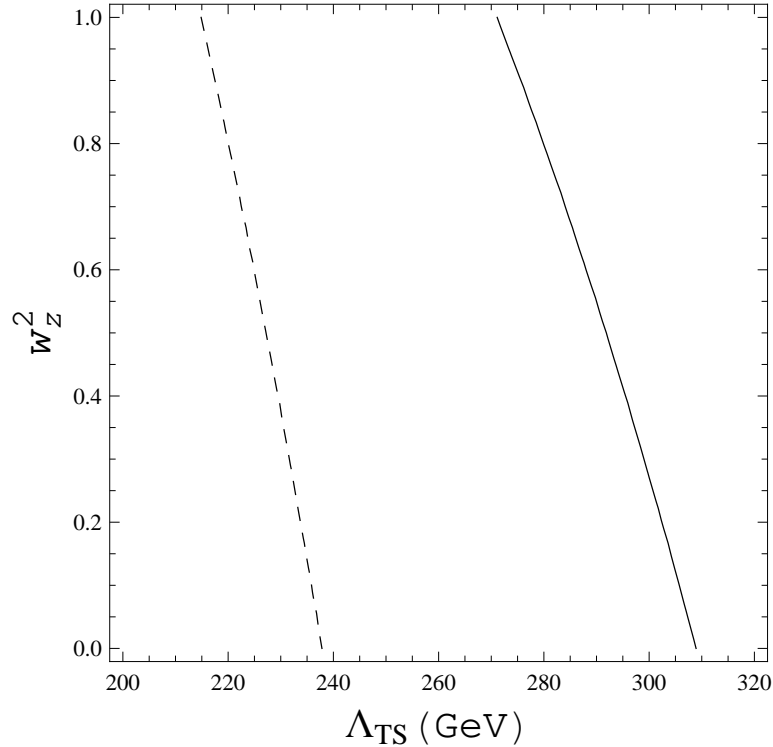}
\caption{
Constraints on parameters $\Lambda_{TS}$ and $w_Z^2$ 
from ALEPH (dashed line) and OPAL (solid line) data.  The left hand side of each of the contours is excluded. 
}
\end{center}
\end{figure}

We finally investigate the implications for phenomenology at the International Linear Collider (ILC), 
with $\sqrt{s}=$500 GeV$-$1 TeV. 
The ILC, with its high precision, can allow us to search 
noncommutative effects for $\Lambda_{TS}$ in the range of $1-10$ TeV. 
Since the total cross section depends on both the location 
of the  ILC and its beam direction, there may be an optimal site and beam direction  for the ILC for observing  noncommutative effects. 
As an example, we take $\Lambda_{TS}$ =500 GeV and 
we calculate the cross section of the process 
$e^+e^- \to q\; {\bar q}$ at the ILC, with $\sqrt{s}=500$ GeV. 
In Figures~2a-c), we show the resulting  deviations of the total cross section 
due to noncommutative effects as a function of 
$\chi$ and $\phi$ for three different values of $w_Z$. (Again, we have cut the region $|\cos \beta| > 0.95$.)
Fig. 2a), where $w_Z^2= 0$, shows that the deviation is maximized for an ILC located at the poles ($\chi=0,\pi$).  On the other hand,
 Fig. 2c), where $w_Z^2= 1$, shows the deviation of the cross 
section is maximized for an ILC located on the equator 
($\chi=\pi/2$) and along the direction to the north ($\phi=0$).  For $w_Z^2= 1$, the deviation tends to zero for an ILC located at the poles, which is evident from the analytic results (\ref{crfrdfsgm}) and (\ref{ttlncsgm}).  Aside from other inconveniences, the poles may therefore not be optimal collider sites for seeing noncommutative effects.

\begin{figure}[t]
\centering \includegraphics[angle=0, width=8cm]{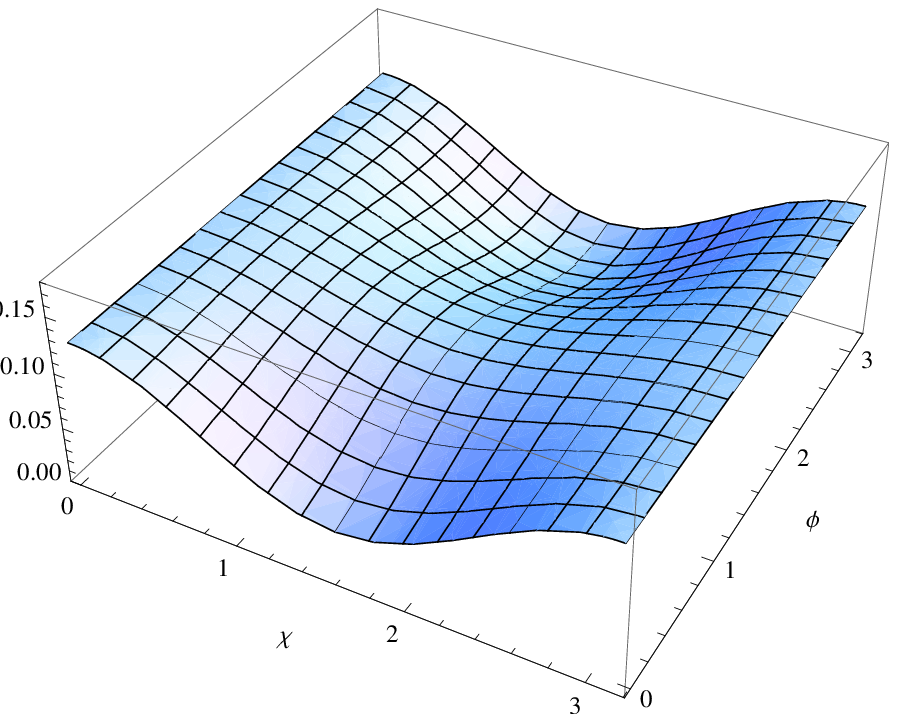}
a)
\centering \includegraphics[angle=0, width=8cm]{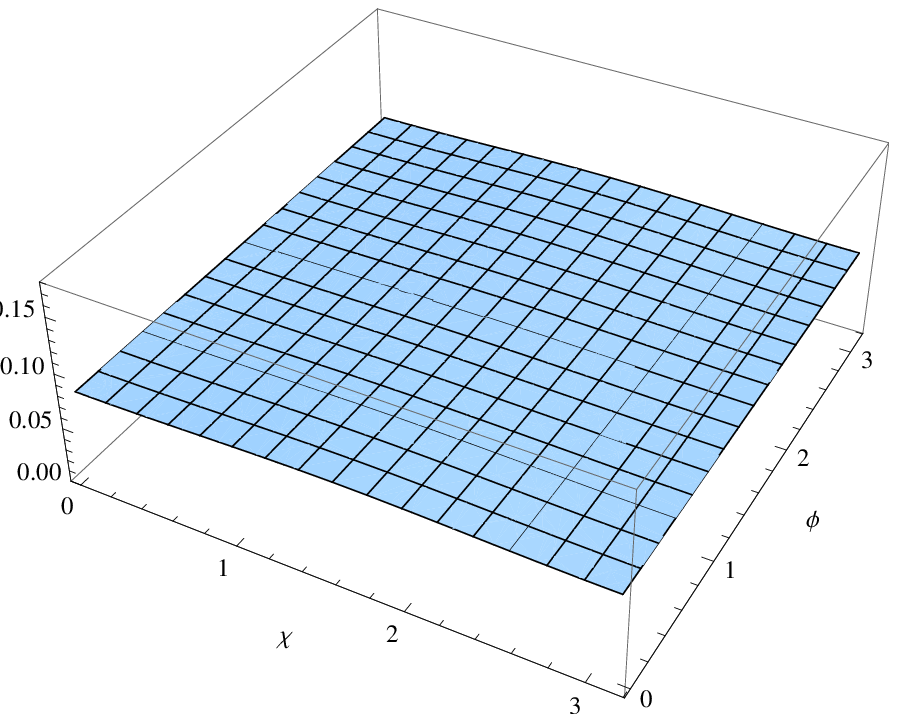}
b)
\centering \includegraphics[angle=0, width=8cm]{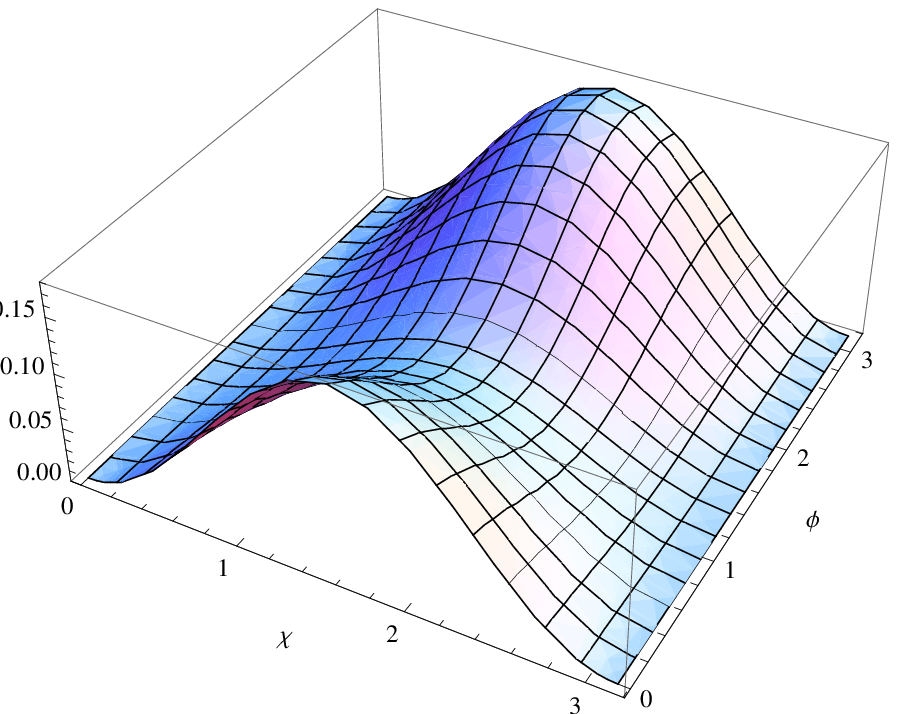}
c)
\caption{
The deviations of the total cross section 
from the standard model one as a function of 
$\chi$ and $\phi$ for a) $w_Z^2=0$, b)
$w_Z^2=0.35$  and c) $w_Z^2=1.0$. 
} 
\end{figure}


\end{document}